\documentclass[namedreferences]{solarphysics}
\usepackage[hyperref,optionalrh,showbiblabels]{spr-sola-addons} 
\usepackage{graphicx}        
\usepackage{epstopdf}
\usepackage{cite}
\usepackage{hyperref}

\usepackage{amssymb}        
\usepackage{color}           

\chardef\us=`\_
\begin{document}

\begin{article}
\begin{opening}

\title{Perception Evaluation -- A new solar image quality metric based on the multi-fractal property of texture features\\ {\it Solar Physics}}

\author[addressref=aff1]{\inits{Y.}\fnm{Yi}~\lnm{Huang}}

\author[addressref={aff1,aff2,aff3},corref,email={robinmartin20@gmail.com}]{\inits{P.}\fnm{Peng}~\lnm{Jia}}

\author[addressref=aff1]{\inits{D.}\fnm{Dongmei}~\lnm{Cai}}
\author[addressref=aff1]{\inits{B.}\fnm{Bojun}~\lnm{Cai}}

\address[id=aff1]{College of Physics and Optoelectronics,
Taiyuan University of Technology, Taiyuan, 030024, China}
\address[id=aff2]{Department of Physics, Durham University, South Road, Durham DH1 3LE, UK}
\address[id=aff3]{Key Laboratory of Advanced Transducers and Intelligent Control Systems, Ministry of Education and Shanxi Province, Taiyuan University of Technology, Taiyuan, 030024, China}
\runningauthor{Yi Huang et al.}
\runningtitle{A new IQM -- PE}
\begin{abstract}
Next-generation ground-based solar observations require good image quality metrics for post-facto processing techniques. Based on the assumption that texture features in solar images are multi-fractal which can be extracted by a trained deep neural network as feature maps, a new reduced-reference objective image quality metric, the perception evaluation is proposed. The perception evaluation is defined as cosine distance of Gram matrix between feature maps extracted from high resolution reference images and that from blurred images. We evaluate performance of the perception evaluation with simulated blurred images and real observation images. The results show that with a high resolution image as reference, the perception evaluation can give robust estimate of image quality for solar images of different scenes. 
\end{abstract}
\end{opening}

\section{Introduction}
     \label{S-Introduction}
The resolution of ground-based telescopes is limited by many different factors such as: the (quasi-) static aberrations or the dynamic aberrations caused by atmospheric turbulence. The atmospheric turbulence induced aberration, termed as ``seeing``, prevents large aperture ground-based solar telescopes from achieving their theoretical angular resolution. For solar telescopes without adaptive optics (AO) systems \citep{thompson2000adaptive}, to alleviate the atmospheric turbulence induced image degradation and achieve higher angular resolution, post-facto image reconstruction techniques are widely used \citep{vanNoort2005,Mikurda2006,Scharmer2010}. A proper objective image quality metric (IQM) is required for these post-facto image reconstruction techniques, because IQM is used either as criterion for frame selection based methods or as cost function for deconvolution algorithms. In recent decades, several IQMs have been proposed and they can be classified into: full-reference (FR), no-reference (NR), and reduced-reference (RR) metrics.\\
FR IQMs require high resolution images as reference. The mean squared error (MSE) is the simplest FR IQM, which computes the average of the squared difference between the distorted and reference image. The structural similarity (SSIM) proposed by \citet{wang2004image} is widely used and it can give similar result as that given by the human visual system. Root-mean-square contrast (RMS-contrast) is the most commonly used IQM for image reconstruction \citep{denker2005high-spatial,denker2007field-dependent,danilovic2008the}. Because the granulation is uniform and isotropic, the FR IQMs has been successfully used for granulation images \citep{Scharmer1989,Denker2005,Danilovic2008}. Unfortunately, FR IQMs have several drawbacks, such as its performance strongly depends on the wavelength \citep{albregtsen1977the} and its sensitivity is related to the structural contents of the image \citep{deng2015objective}.\\
The Median Filter-Gradient Similarity (MFGS) proposed by \citet{deng2015objective} is a NR IQM, which does not need a reference image and is suitable to evaluate the quality of solar observation images directly. However, \citet{popowicz2017review} and \citet{Denker2018} show that MFGS is not completely independent of the structural contents or the spatial sampling rate of an image. This property would limit the performance of image reconstruction methods in different regions of the sun.\\
In this paper, we propose a new RR IQM -- the Perception Evaluation (PE). The PE only requires one high resolution image as reference and can evaluate the quality of blurred images with this reference image. The PE is based on the assumption that texture features in solar images are multi-fractals and they should be similar for solar images obtained in the same wavelength. In this paper, the multi-scale distribution of the multi-fractals is extracted by a trained deep neural network (DNN) \citep{yu2015visual,motoyoshi2007image}. The difference of multi-fractal property between high resolution images and blurred images is then used to evaluate the image quality. We will introduce the PE in Section 2. In Section 3, we will evaluate the performance of PE with simulated blurred images and real observed images and in Section 4, we will give our conclusions and discuss possible applications in the future.\\

\section{Perception Evaluation} 
\label{S-general}
\subsection{Principle of the perception evaluation} 
\label{S-text}
\begin{figure}
\centerline{\includegraphics[width=1\textwidth,clip=]{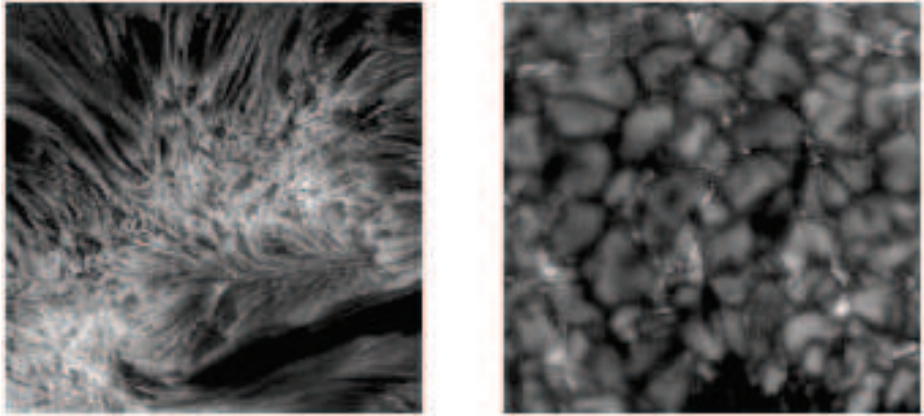}}
          \caption{The high resolution images observed in different wavelengths. On the left is the H-alpha image obtained by the New Vacuum Solar telescope (NVST) and the prominent feature visible there is a dark filament. On the right is the G-band image containing granulation, which is obtained by the Swedish 1-m solar telescope (SST). A skilled astronomer can easily tell the difference between these two figures according to their texture features.}
\label{F-GHimg}
\end{figure}

The texture feature is a description of the spatial arrangement of the gray scale in an image. The texture feature is usually used to describe the regularity or coarseness of an image \citep{guo2012discriminative}. Human beings can easily distinguish between images with different texture features, such as the rainforest or the desert in a black-and-white aerial photograph. For solar images, the texture features are almost everywhere. In different wavelengths, solar images are different and these images are composed of different texture features as shown in Figure~\ref{F-GHimg}.\\
Texture features in solar images may be self-similar and these images are usually called fractal \citep{jia2014parallel}, such as the granulation. In other wavelengths, they are not self-similar in the whole spatial scale, which means they can not be described by a spectrum with the same exponent. However,  these images can be described by a continuous spectrum with different exponents in different scales. This property is usually called multi-fractal property \citep{ne2000the, peng2017discrimination}. If we assume multi-fractal properties of texture features on the solar images do not change between images observed in the same wavelength, with one high resolution image as reference, we can easily discriminate images with different blur level. The difference between texture features of high resolution images and that of blurred images is a good tracer for image quality. Can we model that difference to evaluate image quality?\\
Direct modelling the difference of multi-fractal properties is hard, because texture features are complex and they are different for solar images of different wavelength. Many different image quality metrics based on texture features are proposed, such as: the image grey scale statistics (N-th order joint histograms) introduced by \citet{julesz1962visual}, models based on other statistical measurements \citep{heeger1995pyramid-based,portilla2000a}. Because DNN is complex enough to directly learn texture features, parametric texture feature models based on DNN features are widely used \citep{gatys2015texture,liu2016texture}. In this paper, we will use a trained Convolutional neural network (CNN) to extract the multi-fractal properties of texture features. We will discuss our algorithm below.
\subsection{Algorithm of the perception evaluation} 
\label{S-labels} 
\begin{figure} 
\centerline{\includegraphics[width=1.0\textwidth,clip=]{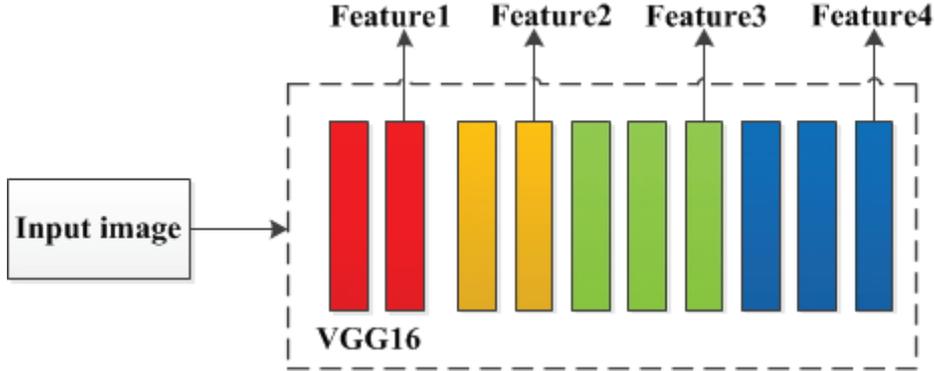}}
          \caption{The structure of the VGG16, which extracts multi-fractal properties from texture features of the input image. A red block represents a convolutional layer with 64 channels and kernel size of 3$\times$3. An orange block represents a convolutional layer with channels of 128 and kernel size of 3$\times$3. A green block and a blue block represents a convolutional layer with channels of 512 and kernel size of 3$\times$3. There is a pooling layer of 2$\times$2 between blocks with different colour. Feature 1, Feature 2, Feature 3 and feature 4 are feature maps of the convolutional layer.}
\label{F-simple1}
\end{figure}

\begin{figure}  
\centerline{\includegraphics[width=1.0\textwidth,clip=]{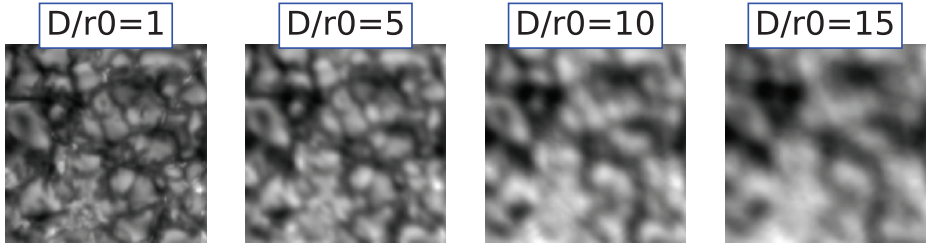}}
          \caption{Short exposure images generated through convolution of Monte Carlo simulated PSFs and high resolution images corresponding to D/r0 = 1,5,10,15.}
\label{F-simple2}
\end{figure}

\begin{figure} 
\centerline{\includegraphics[width=1.0\textwidth,clip=]{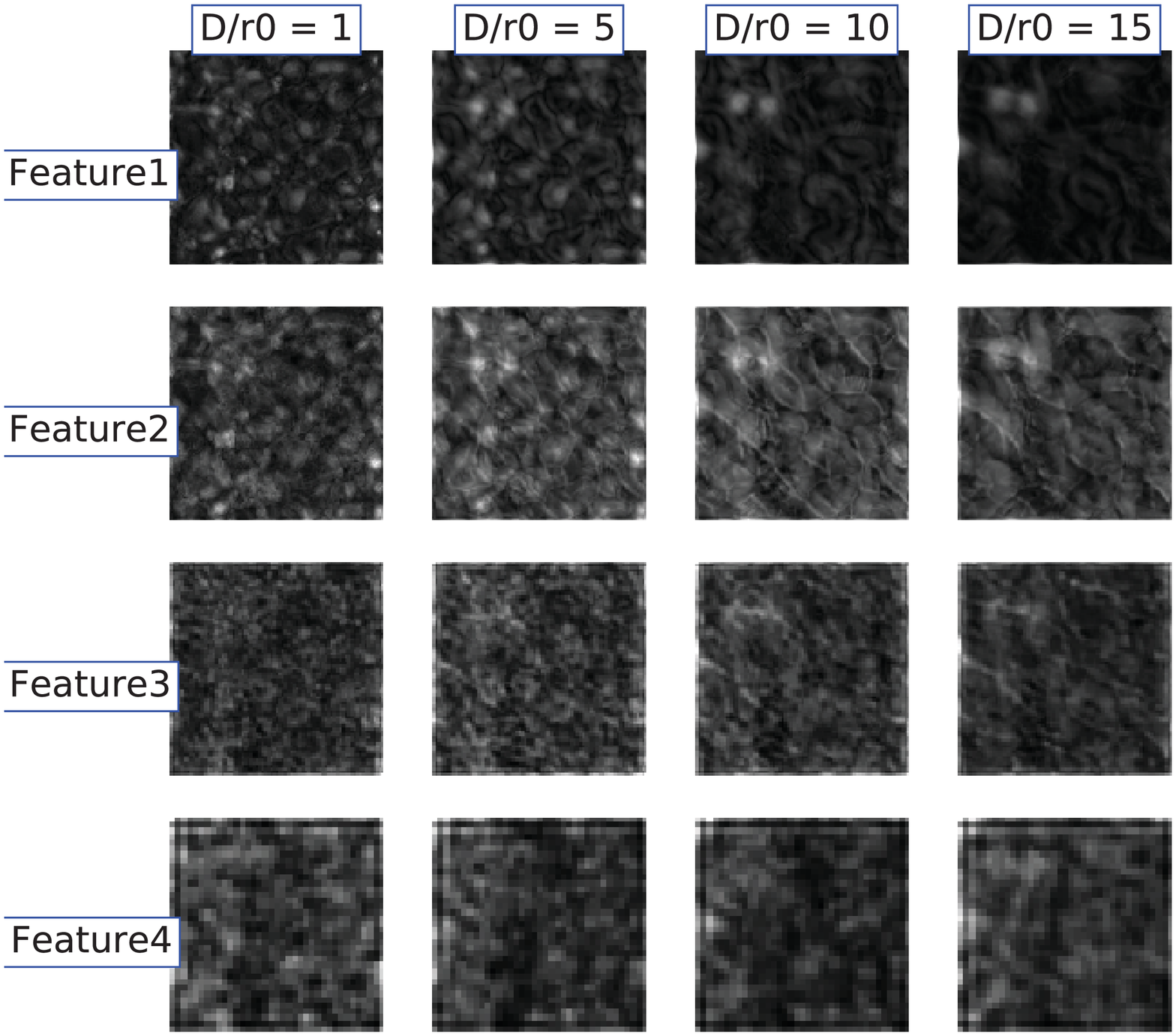}}
          \caption{Images reconstructed from different feature maps (Feature 1, Feature 2, Feature 3, Feature 4) of images with degraded levels of $D/r_0$ = 1,5,10,15.
}
\label{F-simple3}
\end{figure}

CNN is a kind of DNN, which includes many convolutional layers. A convolutional layer has $K$ channels. Different channel means the input signal will be convolved with a different trainable convolutional kernel. The output of each convolutional layers are called feature maps. After a convolutional layer, an image with $M\times N$ pixels will become a 3D feature maps with size of $M\times N \times K$, where $K$ is the channel number. Visualization of feature maps shows that these feature maps describe images in a multi-scale way \citep{zeiler2014visualizing}, which makes it adequate to model multi-fractal properties of texture features on solar images.\\
The VGG, which is a CNN with many small convolution kernels and several convolutional layers and is proposed by the Visual Geometry Group of the University of Oxford \citep{pfister2014deep}, is used in this paper. Considering texture features of solar images are complex and may have different multi-fractal properties, we use VGG16 to model multi-fractal properties of texture features. The VGG16 is a VGG with 12 convolutional layers and 4 pooling layers as shown in Figure~\ref{F-simple1}. In the first several convolutional layers, feature maps from VGG16 have rich details. The deeper the convolutional layer is, the feature maps are more abstract and contain larger scale texture features \citep{gatys2015a}. According to our experience, we select feature maps: Feature 1, Feature 2, Feature 3 and Feature 4 from the trained VGG16 as candidate feature maps in this paper, as shown in Figure~\ref{F-simple1}.\\
To evaluate the representative ability of these feature maps, we generate several short exposure point spread functions (PSF) with different D/r0 through Monte-Carlo simulation \citep{jia2015simulation,basden2018the}. Then we convolve high resolution solar images with these PSFs to generate blurred images as shown in Figure~\ref{F-simple2}. We extract multi-fractal properties from these blurred images by VGG16 and use Feature 1, Feature 2, Feature 3 and Feature 4 to reconstruct these images as shown in Figure~\ref{F-simple3}. These reconstructed images show that feature maps from different layers can reflect multi-fractal properties in different scales. Because image quality metric should be only relevant to blur level, we need to transform feature maps to a quantity that are not relevant to the image size or the structural content.\\
The Gram matrix calculates the correlation between two variables without subtracting their mean values. The Gram matrix can reflect difference between two variables and is normally used for kernel generation for classical machine learning tasks \citep{Hofmann2007}. In recent years, the Gram matrix of feature maps is used in image style transfer to reflect the style difference between two images \citep{johnson2016perceptual} as shown in equation \ref{eq:equation3}.\\
\begin{equation}
\label{eq:equation3}
G_{ij}=\sum_{k}F_{ik}F_{jk}.
\end{equation}
Where $F_{ik}$ and $F_{jk}$ are the 2-dimensional feature maps in a particular layer. To evaluate image quality, the Gram matrix has the following advantages \citep{gatys2015a}:\\
(1) The size of the Gram matrix depends only on the number of feature maps instead of the image size.\\
(2) The Gram matrix is only related to texture features of an image, not its structural content.\\
Thanks to the above advantages, we will use the Gram matrix to represent texture features' multi-fractal properties. In real applications, the Gram matrix of the reference image and that of blurred images will be obtained separately by the VGG16. Then we will calculate the cosine distance \citep{ustyuzhaninov2018one-shot} between these two matrices to evaluate the image quality,
\begin{equation}
\label{eq:equation4}
 L_{quality}=\sum\limits_{i}\sum\limits_{j}\frac{G_{ij}\cdot G^{ref}_{ij}}{\left| G_{ij} \right| \left| G^{ref}_{ij} \right|}.
\end{equation}
where $G_{ij}$ and $G^{ref}_{ij}$ are Gram matrix of the blurred image and that of the reference image, $L_{quality}$ is the PE. According to our experience, the Gram matrix of Feature 4 is best in representing the multi-fractal properties of texture features, because it may contain the largest amount of information which has better expressive ability compared to other feature maps. In VGG16, the size of Feature 4  is $M'\times N' \times 512$, where $M'$ and $N'$ are size of the input signal. In the following sections, we will only use the Gram matrix of Feature 4 to calculate the PE.\\
\section{Performance Evaluation} 
\label{S-features}
\subsection{Sample Data} 
\label{S-equations}
There are two data sets used in this paper: G-band observation data from the SST \citep{scharmer2002dark} (430.5 nm with pixel scale of 0.041 arcsec and exposure time of 4ms) and H-alpha observation data from the NVST \citep{liu2014new} (655.32 nm with pixel scale of 0.136 arcsec and exposure time of 20ms.). The SST data are reconstructed with phase diversity and corrected to the theoretical telescope and detector MTF. The NVST data are reconstructed by speckle reconstruction \citep{li2015high-performance}. All these data are near the diffraction limit and used as reference images in this paper. At the same time, we generate many short exposure PSFs through Monte Carlo simulation \citep{basden2018the}. The parameters in Monte Carlo simulation is set according to Tabel 1 and we will use an accurate atmospheric turbulence phase screen generation method as discussed in \citet{jia2015real-time,jia2015simulation}. These simulated short exposure PSFs will be convolved with reference images to generate simulated blurred images.
\begin{table}
\label{table:1}
\caption{Parameters for Monte Carlo Simulations. The turbulence profile used in this paper is the ESO 35 turbulence profile \citep{Sarazin2013Defining}.}
\begin{tabular}{c c}
\hline\hline
Parameter & Value\\
\hline
Telescope diameter & 980 mm (Gregorian Type) for $H_{\alpha}$ (656.281 nm)\\
                               & 1000 mm  (Schupmann Type) for G-Band (430.5 nm)\\
Atmospheric Turbulence Profile & ESO 35 turbulence profile \\
Fried Parameter ($D/r_0$) & 1 to 20 with step of 1\\
Outer Scale and Inner Scale & 10 m and 0.1 cm\\
Pixel Scale & 0.136 arcsec for $H_{\alpha}$ \\
                  & 0.041 arcsec for G-Band \\
Exposure Time & 20 ms for $H_{\alpha}$\\
    &  4 ms for G-Band\\
\hline
\end{tabular}
\end{table}

\subsection{Performance of the Perception Evaluation } 
\label{S-equations}
\begin{figure}    
\centerline{
          \includegraphics[width=0.4\textwidth,clip=]{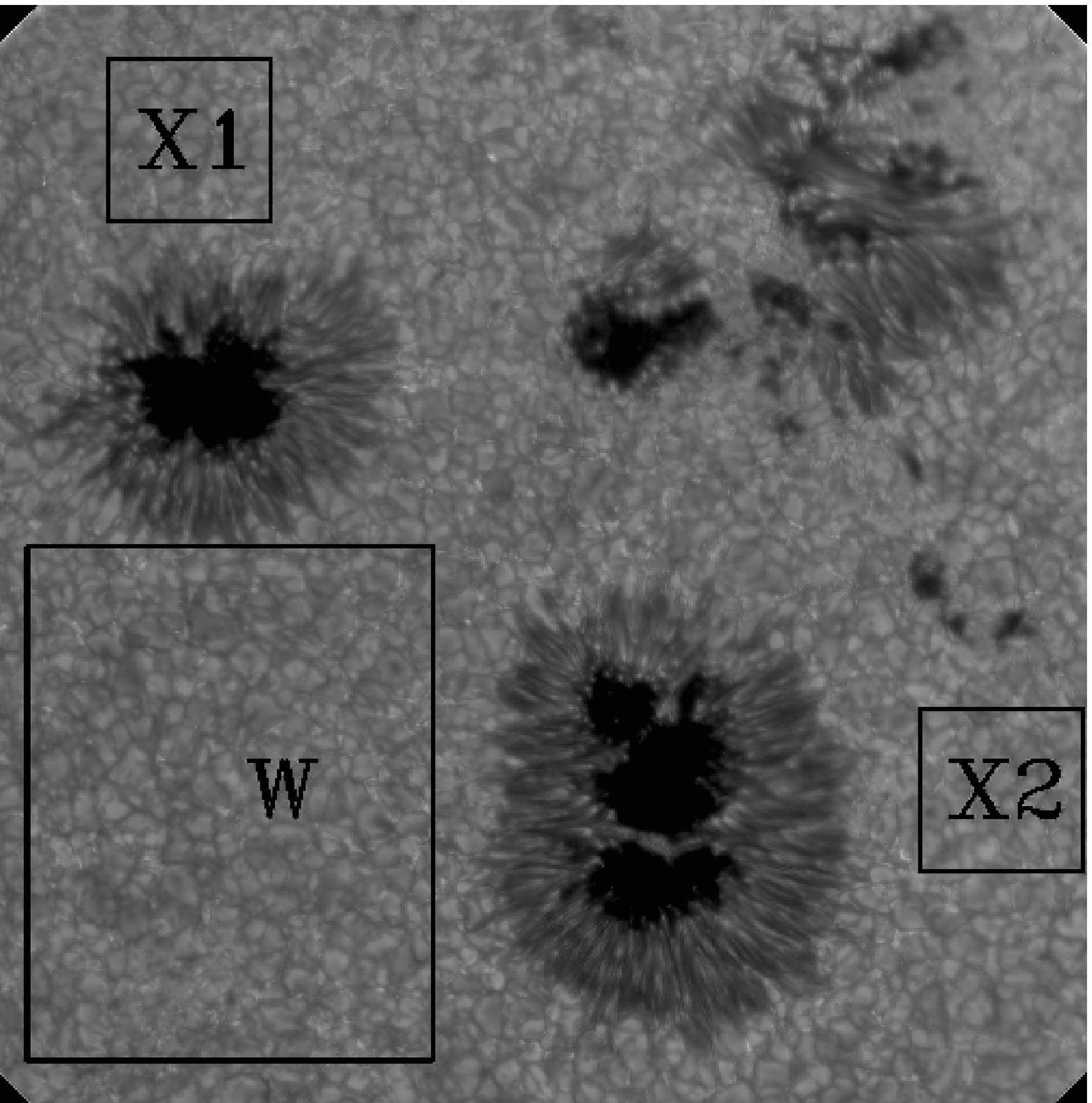}
          \includegraphics[width=0.5\textwidth,clip=]{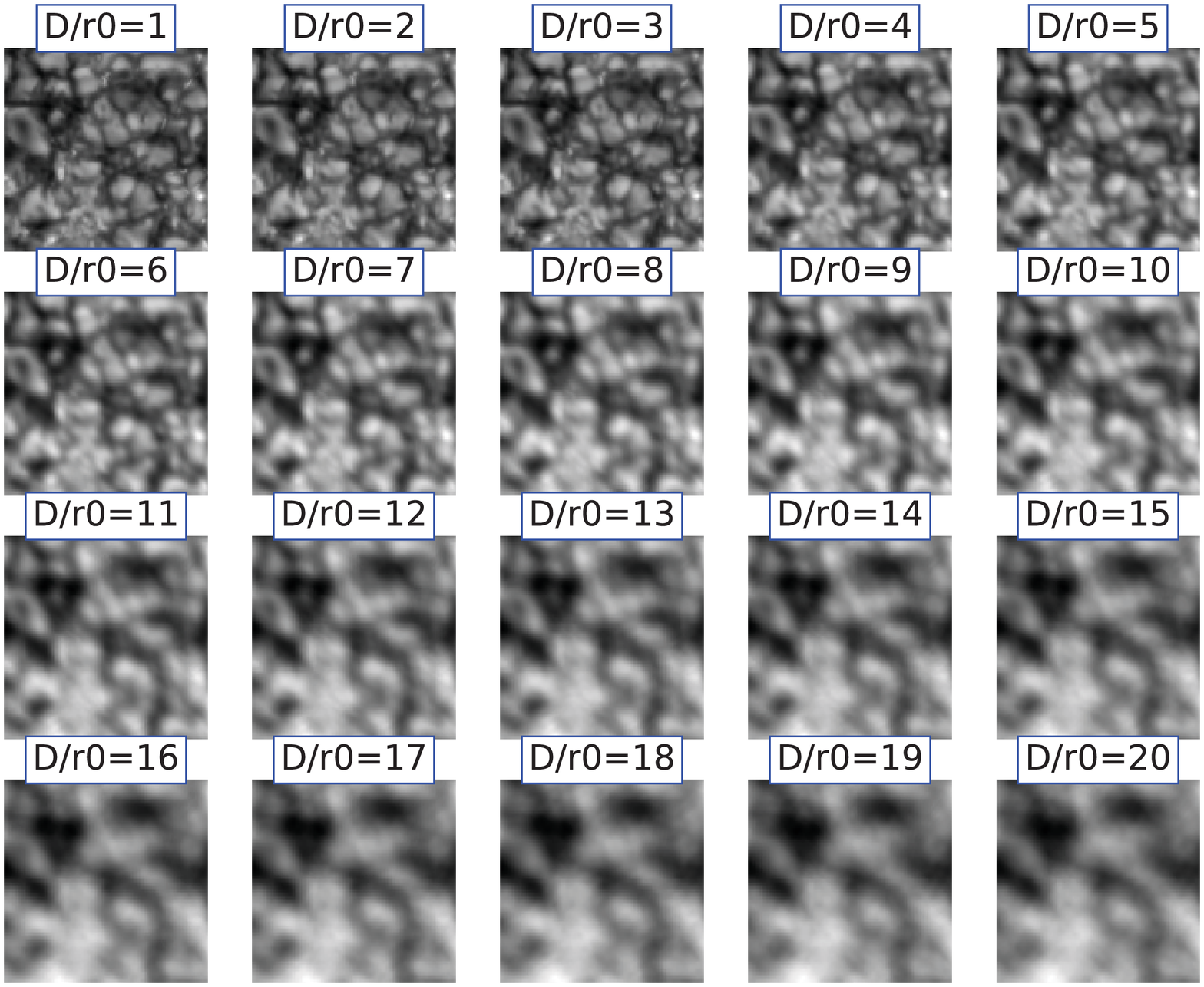}
          }
          \caption{Left panel is the G-band image x1,2 for the selected Standard map area, W as the area to be evaluated, right panel is a patch degradation effect.}
\label{F-simple4}
\end{figure}

\begin{figure}    
\includegraphics[width=1.0\textwidth,clip=]{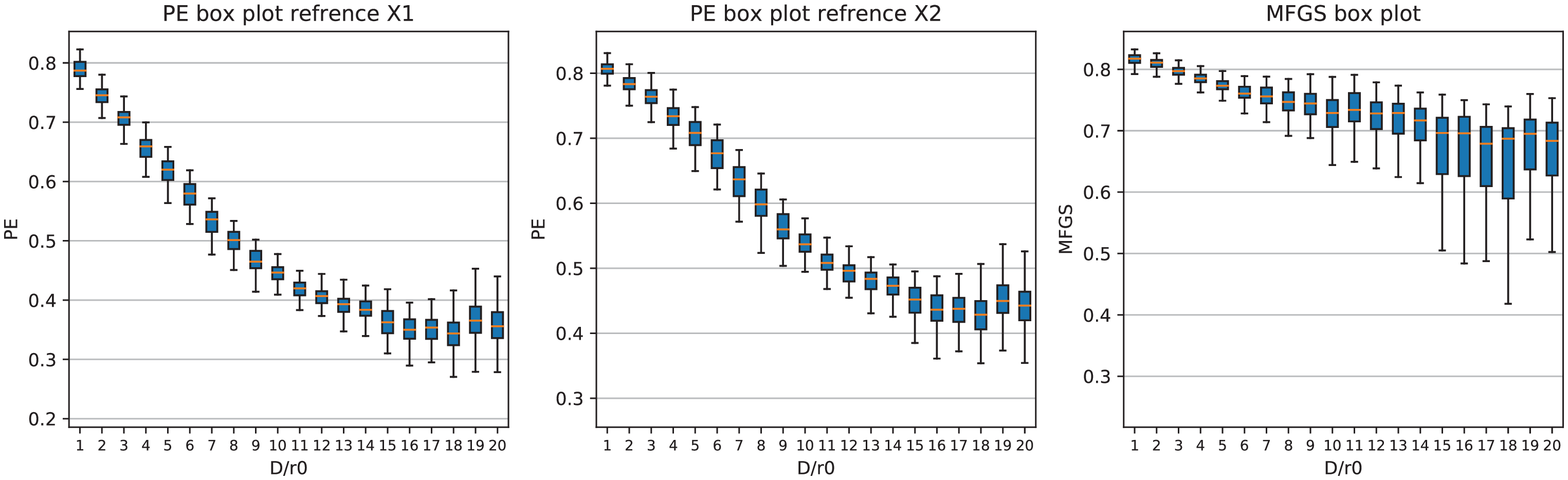}
\caption{Comparison between PE and MFGS with G-band images. The left two figures are box plots of experimental results of PE with reference image of X1 and X2. The right figure is box plot of the experimental results of the MFGS. The standard error of PE is much smaller than that of MFGS and the PE plot is almost the same with different reference images. It is a box plot figure. The size of the box shows the lower quartile and higher quartile of all the data, while the minimum and maximum of all the data is shown in the error bar. All the box plot figures in this paper are plotted in the same way.}
\label{F-simple5}
\end{figure}

\begin{figure}    
\includegraphics[width=0.5\textwidth,clip=]{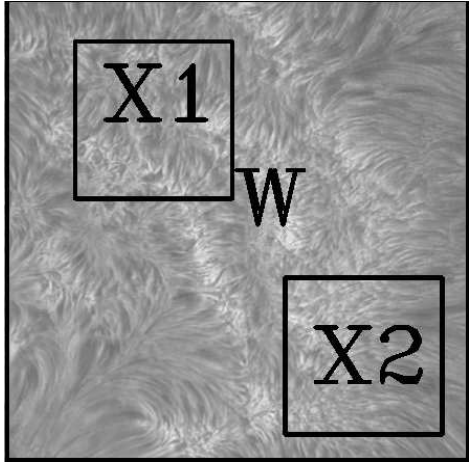}
\caption{The H-alpha data which are obtained by the NVST and we used to test the PE and the MFGS. The regions marked with X1 and X2 are reference images and we extract small images from the region marked with W.}
\label{F6-img_Halpha}
\end{figure}

\begin{figure}    
\includegraphics[width=1.0\textwidth,clip=]{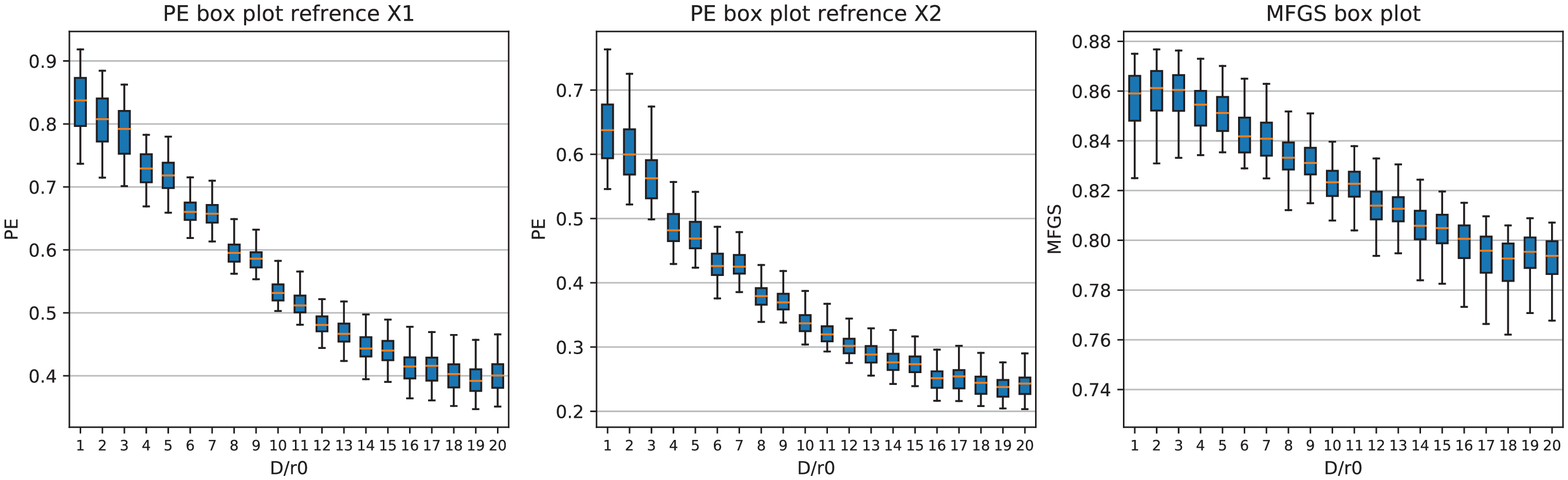}
\caption{Comparison between PE and MFGS with H-alpha images. The left two figures are box plots of experimental results of PE with reference image of X1 and X2. The right figure is box plot of the experimental results of the MFGS. The PE decreases monotonically, when $D/r_0$ is small. The standard error of PE is much smaller than that of MFGS. The trend of PE plot is almost the same with different reference images.}
\label{F-simple6}
\end{figure}

\begin{figure}    
\centerline{
          \includegraphics[width=0.5\textwidth,clip=]{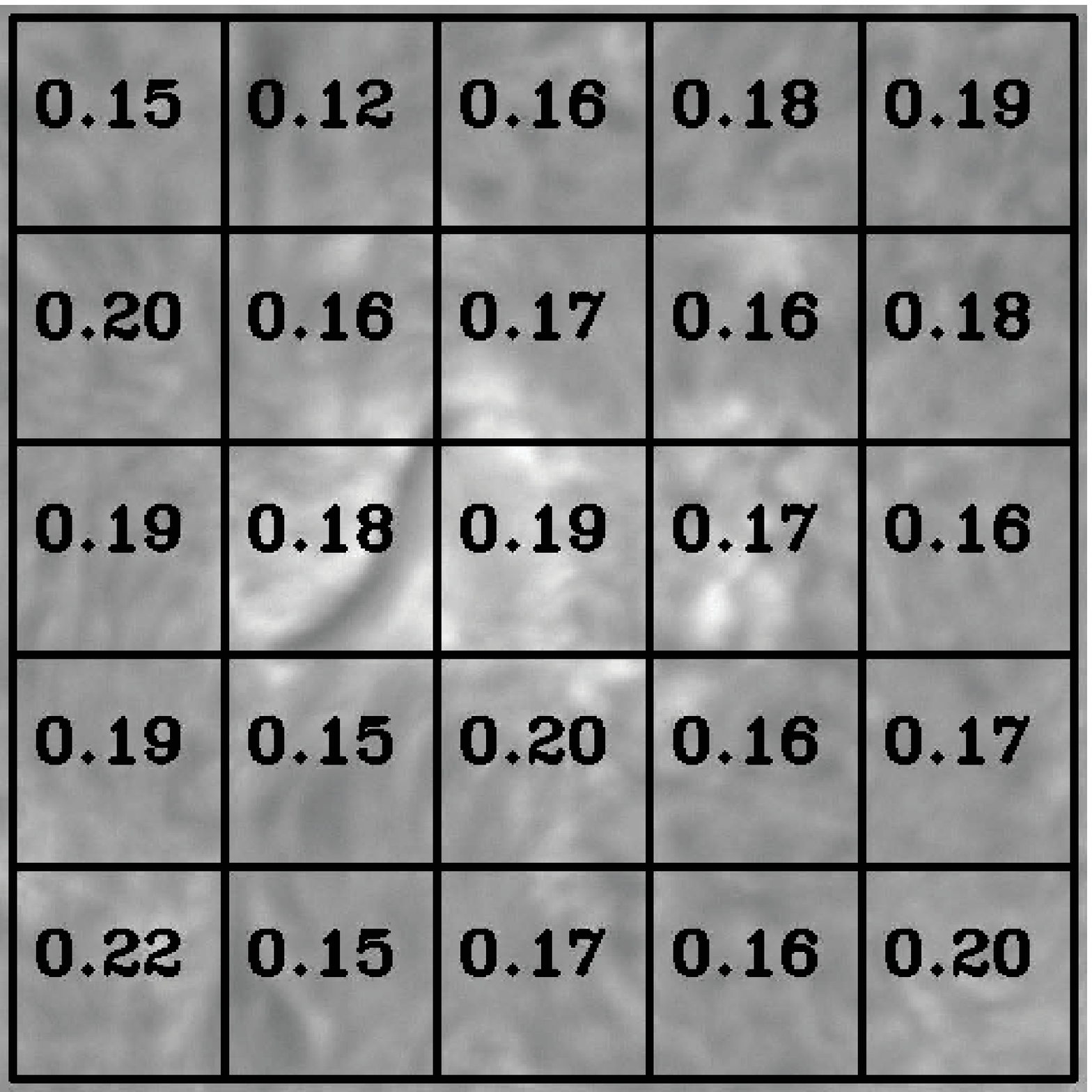}
          \includegraphics[width=0.5\textwidth,clip=]{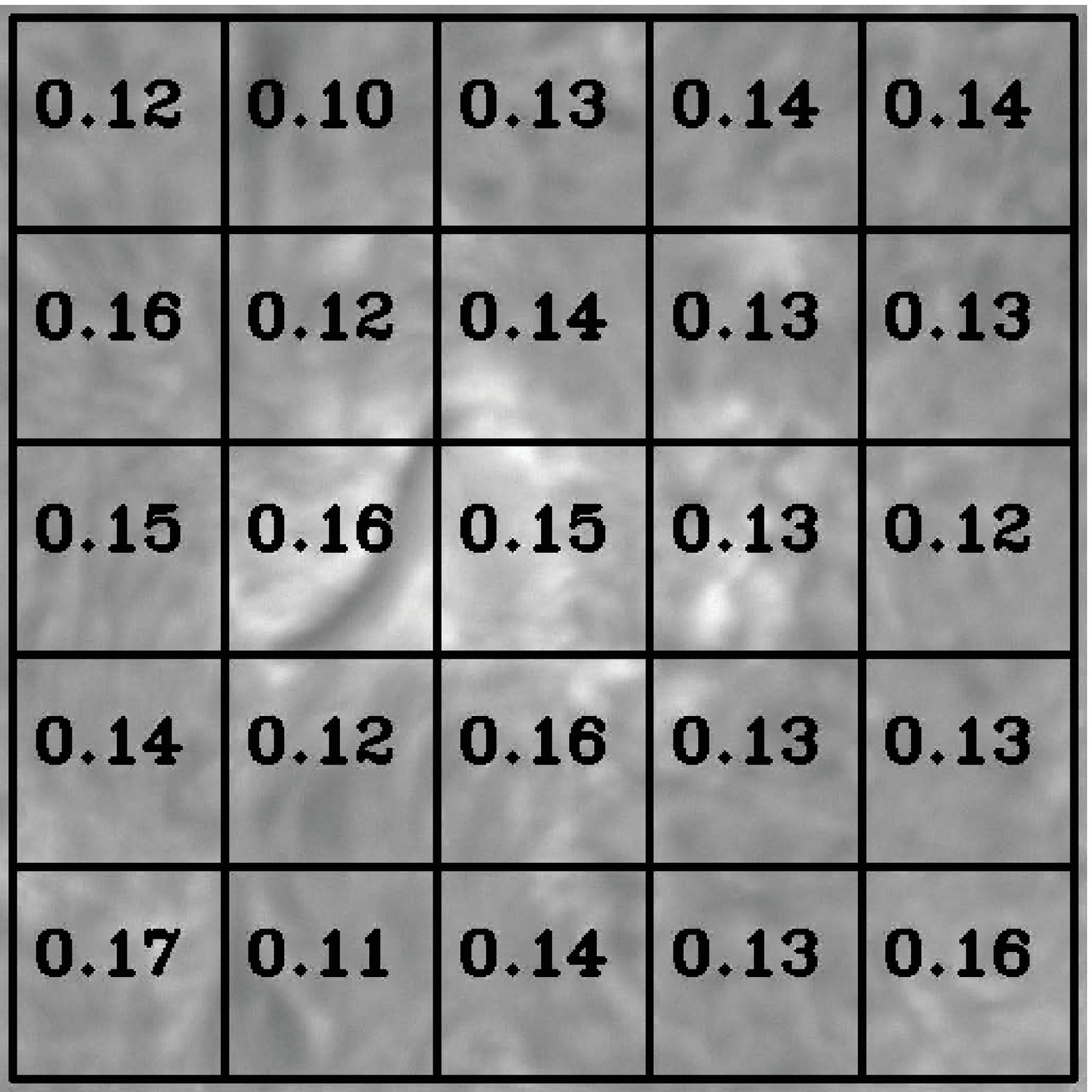}
          }
          \caption{The value of PE in different sections of real observation data. This image is also obtained by the NVST on the same day. The left panel is the PE with X1 as reference and the right panel is the PE with X2 as reference.}
\label{F-simple7}
\end{figure}

\begin{figure}    
          \includegraphics[width=0.7\textwidth,clip=]{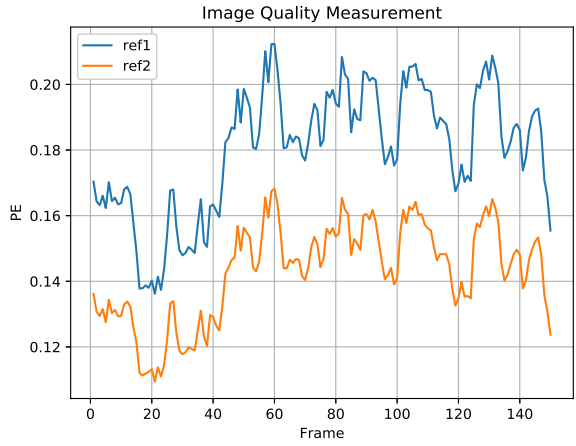}
          \caption{Variation of the mean PE in different frames of real observation H-alpha data. Lines with different colour stand for PE values with different reference images (blue for X1 and orange for X2). As shown in this figure, the absolute value of PE is different when the reference images are different, which may be caused by different amount of texture features in reference images. However, the reference image is large enough to contain all the texture features, because we can find that the PE have the same trend, when reference images are different.}
\label{F-simple8}
\end{figure}

\begin{figure}    
\centerline{\includegraphics[width=1.0\textwidth,clip=]{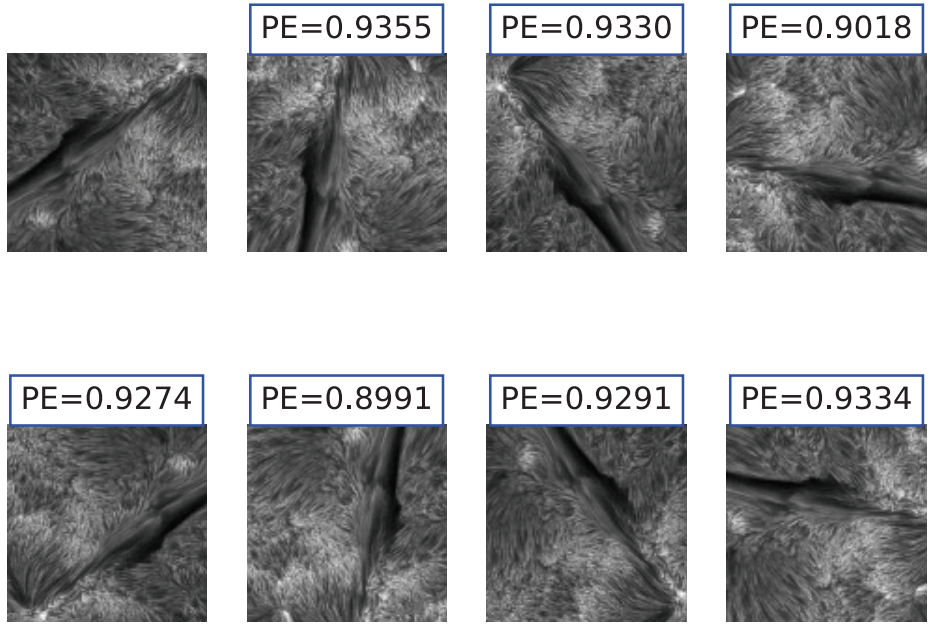}
             }
\caption{Images with different rotation angle and their PE values. The PE values are almost the same for the same images with different rotation angles. }
\label{F7-rotate}
\end{figure}

According to \citet{popowicz2017review}, MFGS proposed by \citet{deng2015objective}, is robust in real applications and considered as a candidate solar image quality metric. In real applications, \citet{Denker2018} has proposed a modified implementation of MFGS to evaluate image sequences obtained with the \textit{High-resolution Fast Imager} (HiFI) at the 1.5-meter GREGOR solar telescope \citep{von2001,Volkmer2010,Denker2012a} and has revealed the field and structure-dependency of MFGS. In this paper we select MFGS for comparison. According to our requriements, we directly add the horizontal and vertical gradients of an image as the MFGS to achieve higher effectiveness.\\ 
Firstly, we use G-band SST observation data as shown in Figure~\ref{F-simple4} to test the PE. Areas X1, X2 with size of $12\times 12$ arcsec are used as reference images. We extract 100 images with size of $300\times 300$ pixels ($12\times 12$ arcsec) from the W area ($750\times 950$  pixels) in Figure~\ref{F-simple4} by step of 50 pixels (around 2 arcsec and these images have overlapping regions). Then we convolve these images with simulated short exposure PSFs (D/r0 from 1 to 20) to generate simulated blurred images. In the right panel of Figure~\ref{F-simple4} are simulated blurred images with different degradation level. These simulated short exposure images are evaluated with PE and MFGS respectively. The results are shown in Figure~\ref{F-simple5}. We can find that PE is more sensitive to different level of blur than MFGS, because the error bar  is much smaller for PE. Besides, we can also find that different reference images will not change the trend of PE, which indicates us that the PE is robust to the reference images.\\
Secondly we use the H-alpha data from NVST to test the PE. As shown in Figure~\ref{F6-img_Halpha}, we extract two reference images X1 and X2 with size of $41 \times 41$ arcsec from H-alpha data as reference images. Then we extract small images with size of $300\times 300$ pixels ($41\times 41$ arcsec) from the W area ($1024\times 1024$ pixels) in Figure~\ref{F6-img_Halpha} by step of 50 pixels (6.8 arcsec) and convolve these images with simulated PSF to generate simulated blurred images. The PE and the MFGS are used to evaluate the quality of these images and the results are shown in Figure~\ref{F-simple6}. We can find that PE still maintains discriminative power for different degrees of image degradation and it is more sensitive than the MFGS.\\
Thirdly, we use PE to evaluate the image quality of real observation data. These real observational images are extracted from the NVST H-alpha observation data on the same day. There are 150 frames of the real observation data and they have size of $1024 \times 1024$ pixels. We use high resolution images X1 and X2 shown in Figure~\ref{F6-img_Halpha} as reference images. One frame of observation images and its PE values in different sections are shown in Figure~\ref{F-simple7}. We can find that the PE can reflect spatial variation of image quality and the variation trend is almost the same for PE with different reference images, which shows that the PE is robust to reference images.\\
Besides, we also evaluate the PE with 150 continuous frames of images. The results are shown in video 1 (see attachment) and Figure \ref{F-simple8}. We can find that the PE can reflect temporal variation of atmospheric turbulence. With different reference images, the absolute value of the PE is different. The difference of the absolute value of the PE is caused by different amount of texture features in different reference images. In real applications, we will use one high resolution image as reference, which means only the relative variation is important. From Figure \ref{F-simple8}, we find that the variation trend of PE is the same when the reference image is different, which indicates the effectiveness of our method.\\
We further explore the stability of PE with the same figure and different rotation angles. We calculate the PE of a speckle reconstructed H-alpha image ($600 \times 600$ pixels) from the NVST with 8 different rotation angles. The reference image is the first image in Figure~\ref{F7-rotate} and as shown in this figure, the difference of PE values between different images are very small, regardless of the rotation angle.\\
\subsection{Limitation of the Perception Evaluation} 
\label{S-Lim}
\begin{figure}    
\centering
         \begin{minipage}{\textwidth}
          \includegraphics[width=1.0\textwidth,clip=]{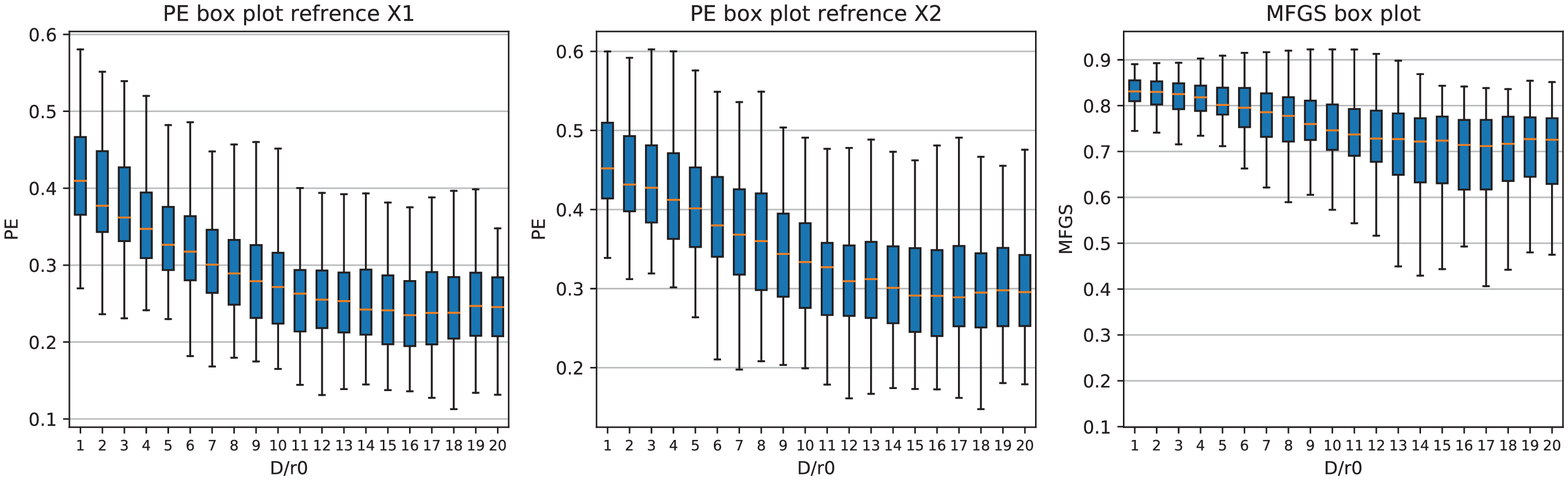}
          \end{minipage}
          
          \begin{minipage}{\textwidth}
          \includegraphics[width=1.0\textwidth,clip=]{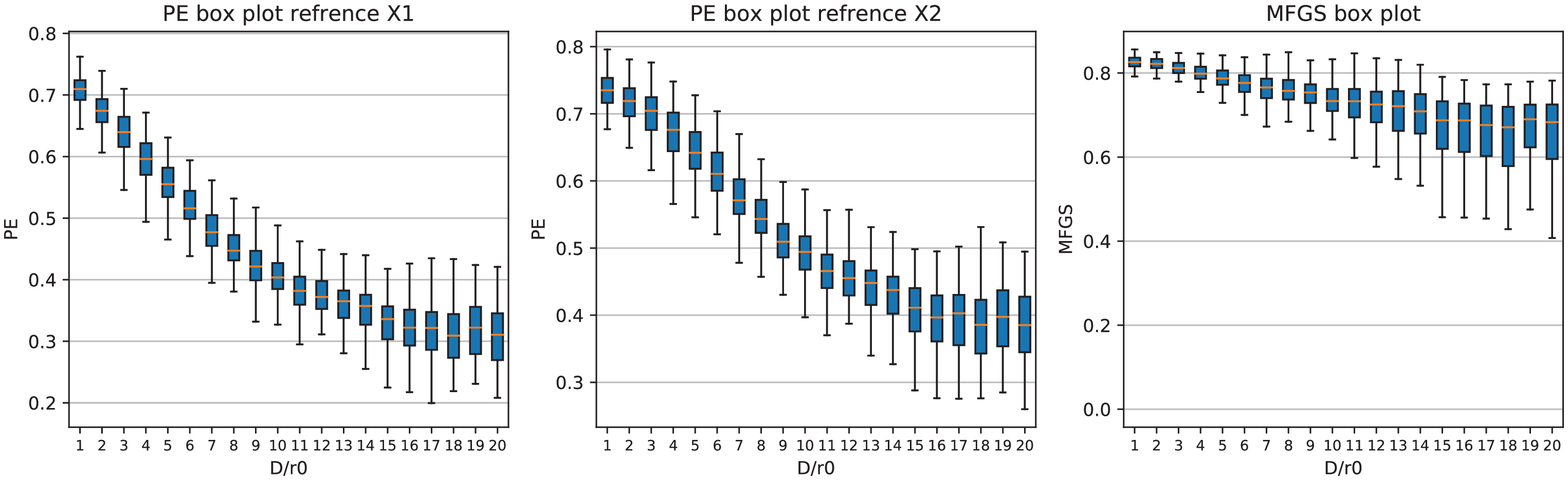}
          \end{minipage}
          
          \begin{minipage}{\textwidth}
          \includegraphics[width=1.0\textwidth,clip=]{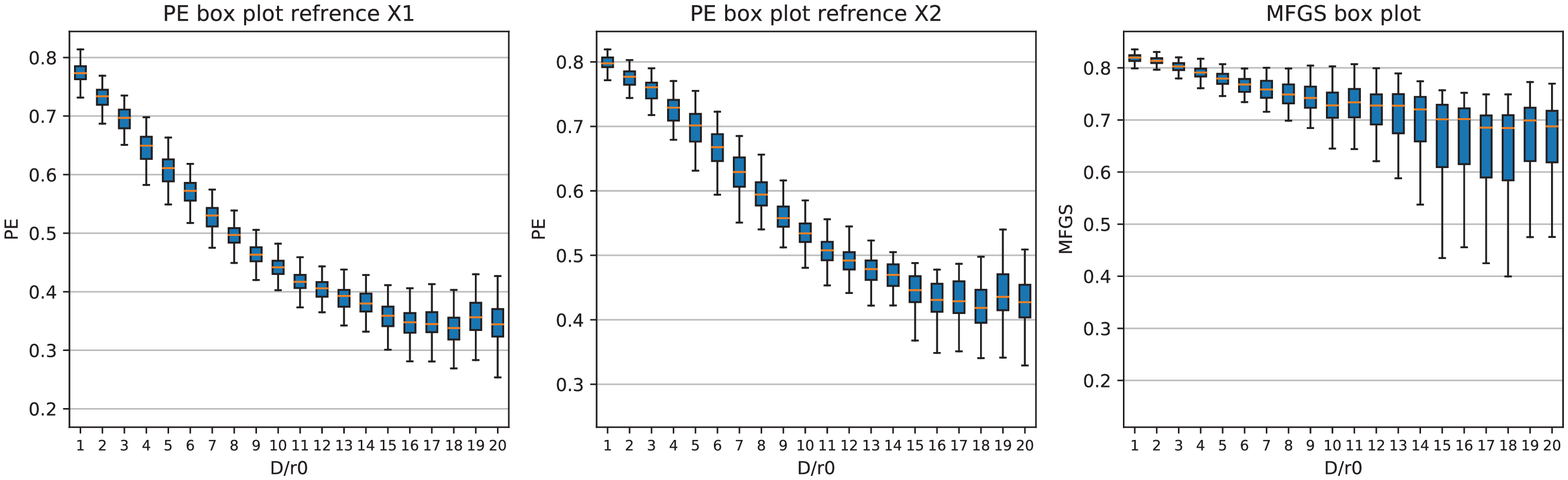}
          \end{minipage}
          \caption{The value of PE and MFGS for G-Band images with different size. The top, the middle and the bottom figures are PE and MFGS value with the same reference image and blurred images of $50\times 50$, $150\times 150$ and $250 \times 250$ pixels respectively.}
\label{F-simple9}
\end{figure}

\begin{figure}    
\centering
         \begin{minipage}{\textwidth}
          \includegraphics[width=1.0\textwidth,clip=]{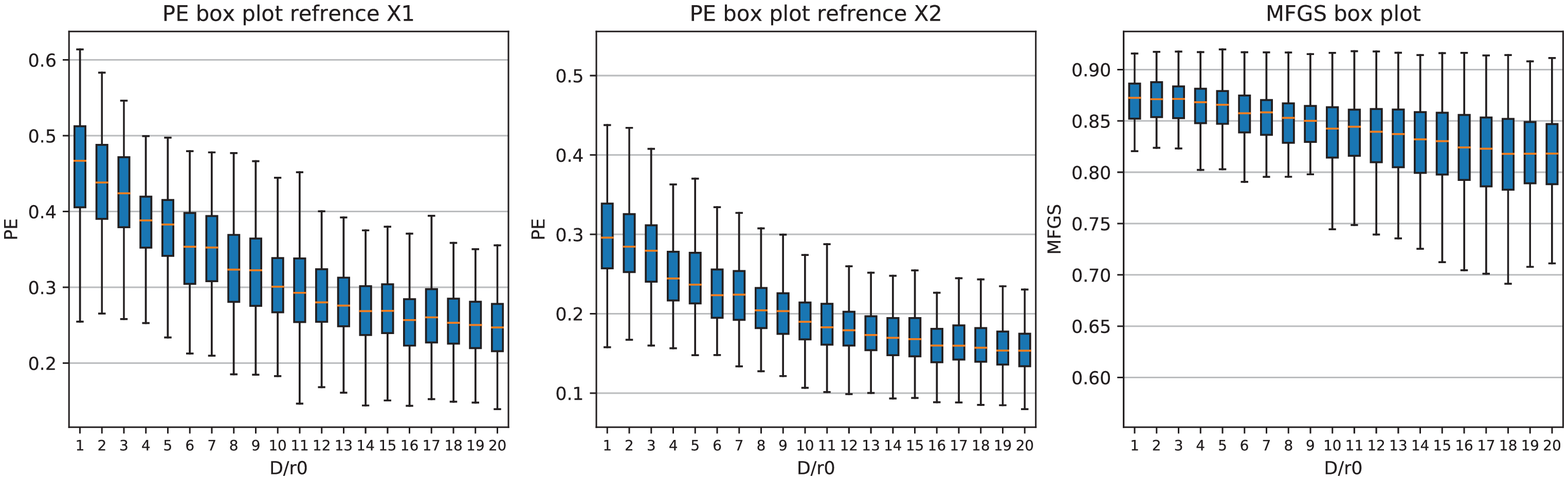}
          \end{minipage}
          
          \begin{minipage}{\textwidth}
          \includegraphics[width=1.0\textwidth,clip=]{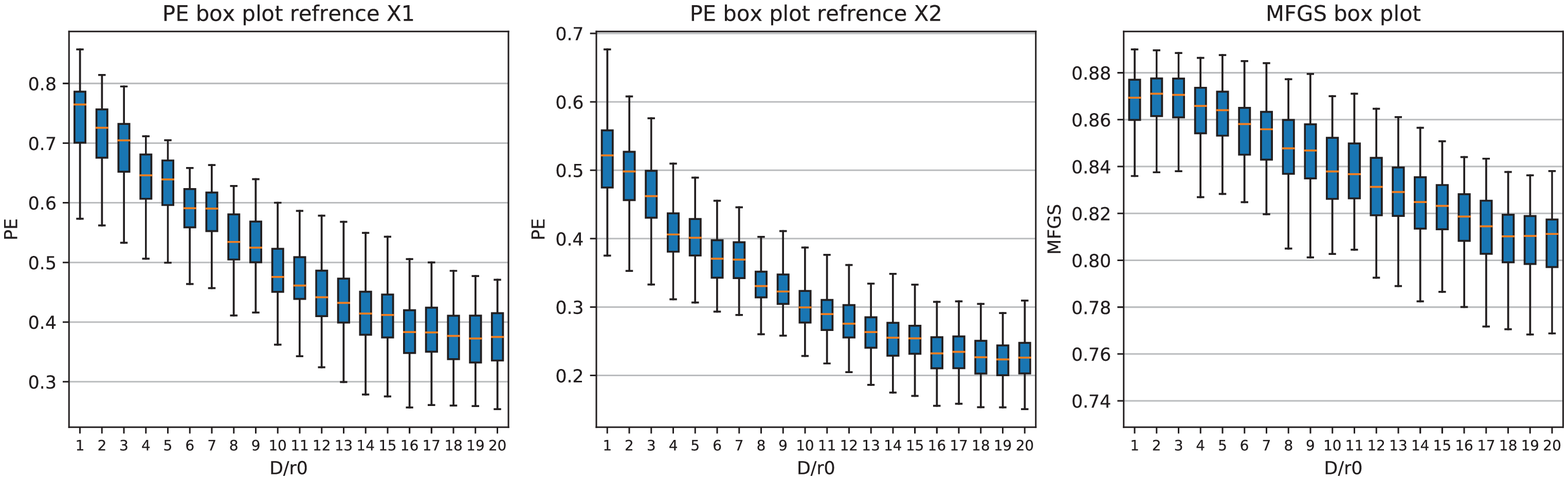}
          \end{minipage}
          
          \begin{minipage}{\textwidth}
          \includegraphics[width=1.0\textwidth,clip=]{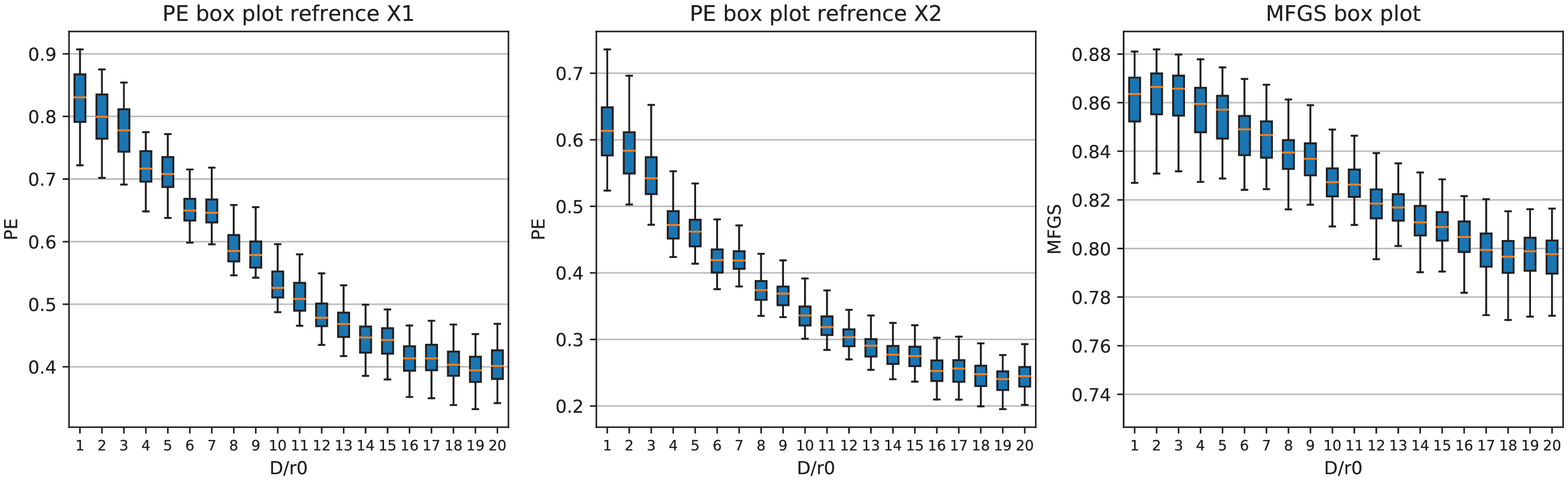}
          \end{minipage}
          \caption{The value of PE and MFGS for H-alpha images with different size. The top, the middle and the bottom figures are PE and MFGS value with the same reference image and blurred images of $50\times 50$, $150\times 150$ and $250 \times 250$ pixels respectively.}
\label{F-simple10}
\end{figure}

We use simulated blurred images with different size and different blur properties (different coherent length) to further test the robustness of the PE and the MFGS. We extract 100 images from W region of Figure~\ref{F-simple4} and Figure~\ref{F6-img_Halpha} and convolve these images with PSFs of different coherent length to generate simulated blurred images. For the PE, we use the same image as reference image for different wavelength (X1 region in Figure~\ref{F-simple4} and Figure~\ref{F6-img_Halpha}). We evaluate the PE and MFGS of these images and the results are shown in Figure~\ref{F-simple9} and Figure~\ref{F-simple10}. We can find that the MFGS and the PE are both sensitive to sampling and image scale. While the PE is sensitive to the image size and it is the limitation of the PE, because the texture features' multi-fractal property is a statistical property and we need a lot of texture features to keep the PE robust. Higher resolution and more pixels in the science camera of future solar telescopes will reduce the limitation of the PE in real applications. Otherwise particular attention should be paid when using PE to evaluate image quality. According to our experiences, images with at least $150 \times 150$ pixels are adequate to be evaluated by the PE.\\

\section{Conclusion} 
\label{S-Conclusion}
Based on the assumption that texture features in the solar image are multi-fractal, we propose a new RR IQMs -- PE in this paper. The PE only needs one high resolution image to evaluate the image quality of blurred images. We test the performance of PE with simulated blurred images and real observation data and find that the PE is robust to image content and rotation angle. However, we also find that the PE is sensitive to the image size. In real applications, we recommend to use PE to evaluate the quality of images which should have at least $150 \times 150$ pixels.\\
Because the PE is robust and only related to texture features of the solar image, we can use it to evaluate the quality of solar images of any wavelength, if we have high resolution image as reference. It will be benefit to frame selection based image restoration methods, because better frames can be selected. The PE can also be directly used as cost function to increase the performance of deconvolution algorithm. Furthermore, the PE can even be used to evaluate the image quality of any astronomy images with texture features, such as nebulae, super nova remnants and galaxies, which would boost up the development of post-facto methods in the astronomical community.\\
\begin{acks}
This work is supported by National Natural Science Foundation of China (NSFC) (11503018) and the Joint Research Fund in Astronomy (U1631133) under cooperative agreement between the NSFC and Chinese Academy of Sciences (CAS), Scientific and Technological Innovation Programs of Higher Education Institutions in Shanxi (2016033). Peng Jia is supported by the China Scholarship Council to study at the University of Durham. The data used in this paper were obtained with the New Vacuum Solar Telescope in Fuxian Solar Observatory of Yunnan Astronomical Observatory, CAS and the Swedish 1-m Solar Telescope. The Swedish 1-m Solar Telescope is operated on the island of La Palma by the Institute for Solar Physics of the Royal Swedish Academy of Sciences in the Spanish Observatorio del Roque de los Muchachos of the Instituto de Astrofísica de Canarias.\\
The authors would like to thank the reviewers for her/his kindly suggestions, which have greatly improved this paper. Peng Jia would like to thank Dr. Yongyuan Xiang, Professor Hui Liu, Professor Kaifan Ji and Professor Zhong Liu from Yunnan Observatory, Professor Hui Deng from Guangzhou University, Dr. Qinming Zhang from Purple Mountain Observatory, Dr. Yang Guo from Nanjing University who provide very helpful suggestions to this paper. The code used in this paper is written in Python programming language (Python Software Foundation) with the package pytroch, astropy and sklearn. The complete code can be downloaded from aojp.lamost.org or https://github.com/yellowyi9527/Perception-Evaluation.\\
\end{acks}

\bibliographystyle{spr-mp-sola}
\bibliography{sola_bibliography_example}
\end{article}
\end{document}